\documentclass[conference]{IEEEtran} 

\usepackage{tikz}
\usepackage{amsmath}

\usepackage{xcolor,color,xspace,enumerate,centernot,multirow,float,graphicx,
xcolor,caption,subcaption,textcomp,pgfplots,pgf-pie,tikz,listings,
comment,adjustbox,mdframed,changepage,algorithm,algorithmic}
\usepackage{hyperref}
\usepackage{filecontents}
\usepackage{xurl}
\usepackage{cryptocode} 
\usetikzlibrary{positioning, shapes.geometric, arrows}

\usepackage{amsmath}
\usepackage{amssymb}
\usepackage{xcolor}
\usepackage{msc}
\usepackage{graphicx}
\usepackage{booktabs}
\usepackage{array}
\usepackage{caption}
\usepackage{cleveref}

\definecolor{gris}{gray}{0.85}

\usepackage[utf8]{inputenc}
\usetikzlibrary{shapes}
\usetikzlibrary{arrows.meta}
\usetikzlibrary{arrows}
\usetikzlibrary{positioning}
\usetikzlibrary{shadows}
\usetikzlibrary{backgrounds}
\usepackage[most]{tcolorbox}
\definecolor{lightblue}{rgb}{0.68, 0.85, 0.9}
\definecolor{lightpink}{rgb}{1.0, 0.71, 0.76}
  {\end{adjustwidth}}

\definecolor{darkgray}{gray}{0.6}

\newmdenv[
  leftline=true,
  rightline=false,
  topline=false,
  bottomline=false,
  linecolor=darkgray,
  linewidth=2pt,
  innerleftmargin=5pt,
  innerrightmargin=0pt,
  innertopmargin=0pt,
  innerbottommargin=0pt
]{leftbar}

\newtcbox{\bluebox}[1][blue]{
  on line,
  colback=blue!5,
  colframe=blue!75!black,
  boxrule=0.1pt,
  arc=1pt,
  boxsep=0pt,left=0.5pt,right=0.5pt,top=0.5pt,bottom=0.5pt
}

\newtcbox{\lightbluebox}[1][blue]{
  on line,
  colback=lightblue,
  colframe=lightblue,
  boxrule=0.1pt,
  arc=1pt,
  boxsep=0pt,left=0.5pt,right=0.5pt,top=0.5pt,bottom=0.5pt
}

\newtcbox{\pinkbox}[1][blue]{
  on line,
  colback=pink!5,
  colframe=pink!75!black,
  boxrule=0.1pt,
  arc=1pt,
  boxsep=0pt,left=0.5pt,right=0.5pt,top=0.5pt,bottom=0.5pt
}

\newtcbox{\lightpinkbox}[1][blue]{
  on line,
  colback=lightpink,
  colframe=lightpink,
  boxrule=0.1pt,
  arc=1pt,
  boxsep=0pt,left=0.5pt,right=0.5pt,top=0.5pt,bottom=0.5pt
}

\AtBeginDocument{%
  }

\newcommand{\vp}[0]{\textsc{Verifpal}\xspace}
\newcommand{\tamarin}[0]{\textsc{Tamarin}\xspace}
\newcommand{\pv}[0]{\textsc{ProVerif}\xspace}

\newcommand{\messLabel}[1]{\small{#1}}

\begin{document}
\title{The Matrix Reloaded: A Mechanized Formal Analysis of the Matrix Cryptographic Suite}

\author{\IEEEauthorblockN{Jacob Ginesin}
\IEEEauthorblockA{Northeastern University}
\and
\IEEEauthorblockN{Cristina Nita-Rotaru}
\IEEEauthorblockA{Northeastern University}}

\maketitle

\begin{abstract}
  Secure instant group messaging applications such as WhatsApp, Facebook Messenger, Matrix, and the Signal Application have become ubiquitous in today's internet, cumulatively serving billions of users. Unlike WhatsApp, for example, Matrix can be deployed in a federated manner, allowing users to choose which server manages their chats. To account for this difference in architecture, Matrix employs two novel cryptographic protocols: Olm, which secures pairwise communications, and Megolm, which relies on Olm and secures group communications. Olm and Megolm are similar to and share security goals with Signal and Sender Keys, which are widely deployed in practice to secure group communications. While Olm, Megolm, and Sender Keys have been manually analyzed in the computational model, no symbolic analysis nor mechanized proofs of correctness exist. Using mechanized proofs and computer-aided analysis is important for cryptographic protocols, as hand-written proofs and analysis are error-prone and often carry subtle mistakes.

Using \pv, we construct formal models of Olm and Megolm, as well as their composition. We prove various properties of interest about Olm and Megolm, including authentication, confidentiality, forward secrecy, and post-compromise security. We also mechanize known limitations, previously discovered attacks, and trivial attacker wins from the specifications and previous literature. Finally, we model Sender Keys and the composition of Signal with Sender Keys in order to draw a comparison with Olm, Megolm, and their composition. Our mechanized analysis indicates subtle yet critical differences in forward secrecy between Olm and Signal, which in turn affects the post-compromise security of Megolm and Sender Keys respectively. 
From our analysis we conclude the composition of Olm and Megolm has comparable security to the composition of Signal and Sender Keys if Olm pre-keys are signed, and provably worse post-compromise security if Olm pre-keys are not signed.

\end{abstract}

\section{Introduction}



End-to-end encryption represents the foundation for security, privacy, trust, and compliance for all services on the internet. Many end-to-end encryption protocols exist, all serving different purposes. Specifically, Transport Layer Security (TLS) \cite{rfc8446} and Quick UDP Internet Connections (QUIC) \cite{rfc9369} secure web traffic, Pretty Good Privacy (PGP) \cite{PGP} secures email communications and files, Zimmermann Real-Time Transport Protocol (ZRTP) \cite{rfc6189} secures Voice over IP (VoIP) communications, while Wireguard \cite{Lipp_Blanchet_Bhargavan_2019} and OpenVPN \cite{OpenVPN} secure point-to-point tunneling. In addition, with the rise of popularity of secure instant messaging applications such as WhatsApp \cite{wb-whatsapp}, iMessage \cite{applePQ3}, Telegram \cite{MTProto}, Facebook Messenger \cite{FBMessenger_2023}, and Matrix \cite{Matrix}, secure instant messaging protocols such as Signal \cite{fm-signal}, OMEMO \cite{OMEMO}, and MTProto \cite{MTProto} have become ubiquitous in today's internet. 

While secure instant messaging protocols have the same high-level guarantees as protocols such as TLS, there are subtleties to instant messaging that lead to differences in protocol design and desired security properties. First, messages are \textit{asynchronous}; an online peer must be able to send messages to an offline peer. This means that parties must rely on a potentially untrusted central server to initiate authentication and key exchange. Second, conversations are \textit{long-lived}; unlike TLS connections, which typically last for a few seconds, instant messaging conversations may go on for years and have many sensitive messages. Thus, it is highly likely an endpoint may be compromised during a conversation lifetime, making the protection of conversation contents in the case of long-term key compromise a major security goal.
Third, transcripts are ideally \textit{restorable}; users often expect to be able to restore their message conversation history from a server, which introduces additional challenges in terms of ensuring the integrity and confidentiality of stored messages.
Thus, besides authentication and confidentiality of messages when long-term keys are not compromised, secure instant messaging protocols are designed to provide security when long-term keys are compromised.
The primary properties of interest
are \textit{forward secrecy}, ensuring \textit{past messages} remain secure in the event long-term keys are compromised, and \textit{post-compromise security}, ensuring \textit{future messages} remain secure in the event long-term keys are compromised. 

Forward secrecy and post-compromise security are well understood for secure two-peer messaging protocols such as Signal \cite{fm-signal}.
However, secure \textit{group} messaging has comparatively seen much less attention \cite{sok_enc_msging_2015}. Group messaging introduces additional complexity, as members join and leave over time, motivating novel protocol designs. One notable protocol for secure instant group messaging is the \textit{Sender Keys} variant of Signal \cite{sender-keys} which, among other applications, is employed by WhatsApp \cite{wb-whatsapp}, Facebook Messenger \cite{FBMessenger_2023}, the Signal Application \cite{signal_app_sk}, and the Session Application \cite{session_blog}.

Matrix is a recently introduced open standard for secure, decentralized, real-time communication that promises interoperability and end-to-end encryption \cite{Matrix}. Unlike Signal, Matrix is designed to be deployed in a federated way, giving users control over which servers host and manage their conversations. Matrix has seen wide adoption across governments, the private sector, and the general public. It is used by government organizations in France, Germany, Sweden, and Luxembourg; for example, France’s central administration takes place on a Matrix-based internal network, Tchap \cite{Tchap}, and Germany’s defense and healthcare organizations employ Matrix in the field \cite{GermanyMatrix}.
Mozilla and KDE use Matrix internally \cite{KDEMatrix}. A number of existing applications integrate Matrix, including the forum software Discourse \cite{RocketChatMatrix} and the enterprise messaging platform Rocket.Chat \cite{RocketChatMatrix}. As of September 2024 Matrix reports over 115 million users on public, data-reporting home servers \cite{MatrixProtocolUsers2023}.

The Matrix stack consists of a number of specifications \cite{Matrix} which cumulatively define a federated secure group messaging protocol. The cumulative protocol allows for users to choose a home server, which in turn federates with other home servers in order to exchange messages. Users construct ”rooms” whose chat history is shared between the users' respective home servers. Every chat in Matrix is a room, including one-on-one chats. The Matrix stack contains three novel protocols: the Matrix protocol, which specifies the various APIs the clients and homeservers must implement, the Olm protocol, which implements peer-to-peer encryption and can be compared to the Signal protocol in terms of security guarantees, and the Megolm protocol, which implements peer-to-multipeer encryption and is comparable to the Sender Keys variant of Signal. The Matrix stack employs TLS to secure individual communications between clients and servers and between servers and other servers (via federation). However, full end-to-end encryption is realized by employing Olm and Megolm in composition. 

%

Since the introduction of Matrix in 2014, several cryptographic vulnerabilities have been reported across both the specifications and implementations of Olm and Megolm. An audit of Olm and Megolm was conducted by the NCC group in 2016 which, among other issues, disclosed an unknown key-share attack in Megolm \cite{ncc-audit}. Several cryptographic vulnerabilities in the Matrix specification and the flagship matrix client, Element, were reported in 2022 \cite{Albrecht_Celi_Dowling_Jones_2023}. Many additional cryptographic vulnerabilities have been reported, including CVE-2021-34813 \cite{CVE-2021-34813}, CVE-2021-40824 \cite{CVE-2021-40824}, CVE-2022-39251 \cite{CVE-2022-39251}, and CVE-2022-39248 \cite{CVE-2022-39248}. Largely in response to \cite{Albrecht_Celi_Dowling_Jones_2023}, an audit of \textit{vodozemac}, a library implementing Olm and Megolm, was conducted by Least Authority. This
audit discovered several implementation-level cryptographic issues \cite{la-audit}. Implementation-level weaknesses were discovered in \textit{libolm}, a recently depreciated yet widely used reference implementation of Olm as of August 2024 \cite{soatok2024gist}. These vulnerabilities highlight the need for precise, formal, and rigorous cryptographic analysis, especially to understand precisely how weaknesses in Olm or Megolm may affect the entire Matrix stack.


Formal security proofs of cryptographic protocols and cryptosystems are generally classified into two categories: (1) \textit{computational} proofs, which make precise computational assumptions and show the complexity of breaking the cryptosystem reduces to the complexity of breaking the assumptions; (2) \textit{symbolic} or Dolev-Yao proofs, which rely on algebraic abstractions of primitives and directly reason about the actions of a channel-controlling attacker. In general, both methods can be used to provide complementary benefits as some attacks may arise from one approach but not the other \cite{nadim-automated}. A full breakdown of the differences between the two approaches is discussed in greater detail in \cite{Blanchet2012}. In recent years it has become the standard to analyze cryptographic protocols both symbolically and computationally, as demonstrated by the many works on Signal \cite{nadim-automated} \cite{fm-signal} \cite{fm-signal-cm} \cite{Bhargavan_dy} \cite{Bhargavan_Jacomme_Kiefer_Schmidt} \cite{vp}, TLS \cite{Cremer_2017}, and Wireguard \cite{Donenfeld} \cite{Dowling_Paterson} \cite{Lipp} \cite{Kobeissi_Bhargavan}, as well as the dual analysis of iMessage PQ3 \cite{Stebila} \cite{Basin_Linker_Sasse}. 


\textbf{Our contribution.} In this work we take an approach rooted in \textit{formal methods} and \textit{computer-aided reasoning} to study the cryptographic security of the Matrix protocol suite, allowing us to construct computer-verified proofs or verified and explicit counterexamples.  We employ \pv as our symbolic analysis tools of choice. We make the following contributions:


\textbf{Models}. We present the first \textit{symbolic} analysis and mechanization of the Matrix protocol suite; taking such an approach allows us to precisely and automatically compare the cryptographic guarantees of the Matrix cryptographic suite to its contemporaries. We construct the first formal models for Olm, Megolm, 
and their composition. To understand how Olm and Megolm compare with Signal and
Sender Keys, we employ a pre-existing \pv Signal model \cite{nadim-automated}, we implement Sender Keys, and we model the composition of Signal and Sender Keys. 

\textbf{Verification}. Using \pv for automated symbolic analysis, we derive various proofs, limitations, and previously found attacks from the Matrix and Sender Keys specifications and previous formalizations. First, we mechanically prove the confidentiality, authentication, forward secrecy, and post-compromise security properties of Olm and Megolm, mechanizing the previous hand-written computational analysis of the Matrix cryptographic core \cite{Albrecht2024} and the stated security goals in the Olm \cite{olm-docs} and Megolm \cite{megolm-docs} specifications. We then mechanize previously known attacks and limitations, including the Megolm unknown key-share attack described in the Matrix NCC audit \cite{ncc-audit}, the tradeoff between forward secrecy and deniability in Olm \cite{olm-docs}, and the unauthorized injection attack described in the computational analysis of Sender Keys \cite{Balbas_Collins_Gajland_2023}. We mechanically prove the extension of Sender Keys proposed by Balbas et al. satisfies the intended security properties, confidentiality and authentication \cite{Balbas_Collins_Gajland_2023}. 

\textbf{Comparison}. Using our \pv models, we precisely compare the cryptographic guarantees of Olm and Megolm with Signal and Sender keys. Among other insights, we observe Olm channels without pre-key signing result in the post-compromise security of the composition of Megolm being significantly weaker than Sender Keys. 
We note that the Megolm specification \cite{megolm-docs}
does not require Olm pre-keys to be signed while the Matrix protocol \cite{matrix-spec} requires the Olm pre-keys to be signed but provides no reason why. Our work elucidates this question providing formal proofs that in the absence of Olm pre-keys signed, the composition of Olm and Megolm provides weaker security.


\textbf{Ethics consideration.}
We have alerted the Matrix foundation's security contact point of our findings and risks. They pointed us to the design
document specifiying the Matrix architecture \cite{matrix-spec} that require Olm pre-keys to be pre-signed, thus confirming our findings. 

\textbf{Code}. Our models are available in our anonymized git repository \url{https://anonymous.4open.science/r/Matrix-Symbolic-Analysis-7D9D}.

\section{Background}


In this section, we present the necessary background for our work, including a description of Signal, Sender Keys, Olm, Megolm, and a brief overview of their security. We then clarify the scope of our analysis.

\subsection{Overview of Signal and Sender Keys}
\textbf{Signal}. The Signal protocol is a cryptographic protocol used for encrypted instant messaging between two parties. Among other properties, Signal provides confidentiality, integrity, forward secrecy, and post-compromise security \cite{sok_enc_msging_2015}.  Signal provides end-to-end encryption for Whatsapp \cite{wb-whatsapp} and Facebook Messenger \cite{FBMessenger_2023} among many other services, serving over a billion users \cite{fm-signal}. 

The core of Signal's design is the double ratchet algorithm, which provides forward secrecy and post-compromise security simultaneously \cite{fm-signal}. Signal connections are chunked into Signal "sessions," which indicate stateful cryptographic context in which forward secrecy is held between. The symmetric key for each (short-lived) Signal session is derived from a combination of (1) a ratchet, defined as a key which is continually "advanced" via a hashing algorithm per each new Signal session, agreed upon at the beginning of the Signal connection, and (2) a "fresh" shared key derived from a Diffie-Hellman key exchange, which both peers take turns in initiating. To securely initiate an authenticated connection and agree upon a shared initial ratchet key, Signal relies on the Extended Triple Diffie-Hellman (X3DH) key agreement protocol \cite{X3DH}. 

In more detail, the protocol works as follows. Each peer generates an asymmetric long-term identity key pair, a medium-term asymmetric pre-key pair, and a one-time asymmetric pre-key pair. Then, each peer uploads a "pre-key bundle" to the Signal server, containing their public identity key, their public pre-key signed with the identity key, and the public one-time pre-key. If Alice wanted to initiate a connection with Bob, Alice would retrieve Bob's pre-key bundle from the Signal server, generate an asymmetric ephemeral key pair, and compute four Diffie-Hellman key agreements to derive the shared initial master key: (1) between Alice's ephemeral key and Bob's identity key, (2) between Alice's identity key and Bob's signed pre-key, (3) between Alice's ephemeral key and Bob's signed pre-key, and (4) between Alice's ephemeral key and Bob's one-time pre-key. Alice then sends Bob her ephemeral public key, her identity key, and some additional metadata; Bob then computes the Diffie-Hellman exchanges in turn, deriving the same master key as Alice. 
Once Alice and Bob agree on a shared master key, they derive an initial ratchet key using a key derivation function. 
Importantly, this key exchange provides post-compromise security in the case long-term identity keys are leaked. However, Signal is reliant on off-band verification of identity keys \cite{fm-signal}, and without such the protocol is vulnerable to an unknown key-share attack. 

The Signal handshake is visualized in Figure \ref{fig:signal} in the Appendix, and we refer the reader to \cite{fm-signal} for a complete formal description of the protocol.

  \textbf{Sender Keys}. The Sender Keys protocol is a cryptographic protocol used for encrypted instant messaging within a group \textit{without} group key agreement. In comparison to protocols with group key agreement such as Message Layer Security, Sender Keys offers different performance guarantees, as fully detailed in \cite{Bhargavan_Beurdouche_Naldurg}. Sender Keys provides confidentiality, forward secrecy, and authentication \textit{assuming} there exists a secure peer-to-peer channel (such as Signal). However, because Sender Keys does not do group key agreement, \textit{transcript equivalence} is not maintained \cite{rfc9420}. That is, the message histories between Sender Keys peers may differ due to asynchronous message delivery or the re-ordering of messages. Sender Keys has seen widespread deployment, being used by the Signal Application \cite{signal_app_sk}, Whatsapp \cite{wb-whatsapp}, Facebook Messenger \cite{FBMessenger_2023}, and the Session Application \cite{session_blog}. 

  The core mechanism of the Sender Keys protocol is the notion of a \textit{session}. Note, sessions in the context of Sender Keys or Megolm are entirely different from sessions in Signal or Olm; notionally, we will prefix each mention of a session with the associated protocol for the remainder of the paper. 

  The protocol works as follows. Each time the state of the group changes (e.g., a member joins, leaves, etc.), each peer generates a new Sender Keys session and transmits it to all other peers in the group via a secure peer-to-peer channel. A Sender Keys session consists of an asymmetric "sender" keypair (of which only the public key is transmitted), a ratchet key, and a ratchet index. Sent messages are signed with the private key of the sender keypair and encrypted with a symmetric key derived from the ratchet key, which is advanced as necessary with a hash function. All receiving peers verify the signature with the public key from the sender keypair, advance the ratchet key, derive the symmetric key, and decrypt the message. Thus, each peer stores the Sender Keys sessions of all other peers. A full Sender Keys exchange is defined in Figure \ref{fig:megolm-sk}.

\subsection{Overview of Olm and Megolm}%
\label{sub:Olm and Megolm}

While the Signal and Sender Keys protocol scheme assumes a single, reliable server, Matrix is a \textit{federated} system with no central server. This motivated the development of Olm and Megolm (adjacent to the development of Signal and Sender Keys), cryptographic protocols that are better designed to be routed between multiple servers as defined via the Matrix protocol's federation scheme \cite{Matrix}. The Matrix protocol specification, which defines room construction and message routing between federated servers and employs Olm and Megolm as sub-protocols, is disjoint from the protocol specifications of Olm and Megolm. For the purpose of our analysis we do not reason about the overarching Matrix protocol and instead reason directly about Olm and Megolm.



\textbf{Olm}. Olm is used in one-on-one chats between two parties, and can be compared to the Signal protocol in terms of security guarantees \cite{fm-signal} \cite{olm-docs}. However, Olm is never used individually, and its purpose is to serve as a subroutine for initial key exchange for Megolm. Precisely like Signal, Olm is a double ratchet algorithm that derives each Olm session key from advancing a ratchet key and a deriving a shared key via Diffie-Hellman (which, like Signal, both peers take turns in initiating). Where Olm differs from Signal, however, is the algorithm for which Olm initiates connections: Olm utilizes Triple Diffie-Hellman (3DH), while Signal utilizes Extended Triple Diffie-Hellman (X3DH). 

The protocol works as follows. Each peer generates an asymmetric long-term identity key pair and an asymmetric one-time key-pair. 
Then, each peer uploads their pre-key bundle to the Olm server, consisting only of the peer's identity key and public one-time key. If Alice wants to initiate a connection with Bob, Alice would retrieve Bob's pre-key bundle from the Olm server and compute three Diffie-Hellman key agreements to derive the initial shared ratchet key:
(1) between Alice's identity key and Bob's one-time pre-key, (2) between Alice's one-time pre-key and Bob's identity key, and (3) between Alice's one-time pre-key and Bob's one-time pre-key. 

A full breakdown of the differences between Olm and Megolm is available in Table \ref{tab:p2p_comparison}. Most notably, Olm \textit{does not require} one-time pre-keys be signed, citing a tradeoff between deniability and forward secrecy \cite{olm-docs}. Additionally, because Olm uses Curve25519 as opposed to Signal's XEdDSA scheme, Olm must generate its own Fingerprint key -- a short sequence of bytes used to identify a longer public key -- as opposed to deriving it from the identity key as Signal does \cite{Perrin}. 
We note that the Matrix protocol specification diverges from Olm \cite{olm-docs} and \textit{requires} Olm pre-keys to be signed \cite{matrix2024client}, yet there does not exist any justification for this decision.  


\definecolor{gris}{gray}{0.85}
\begin{figure*}[ht!]
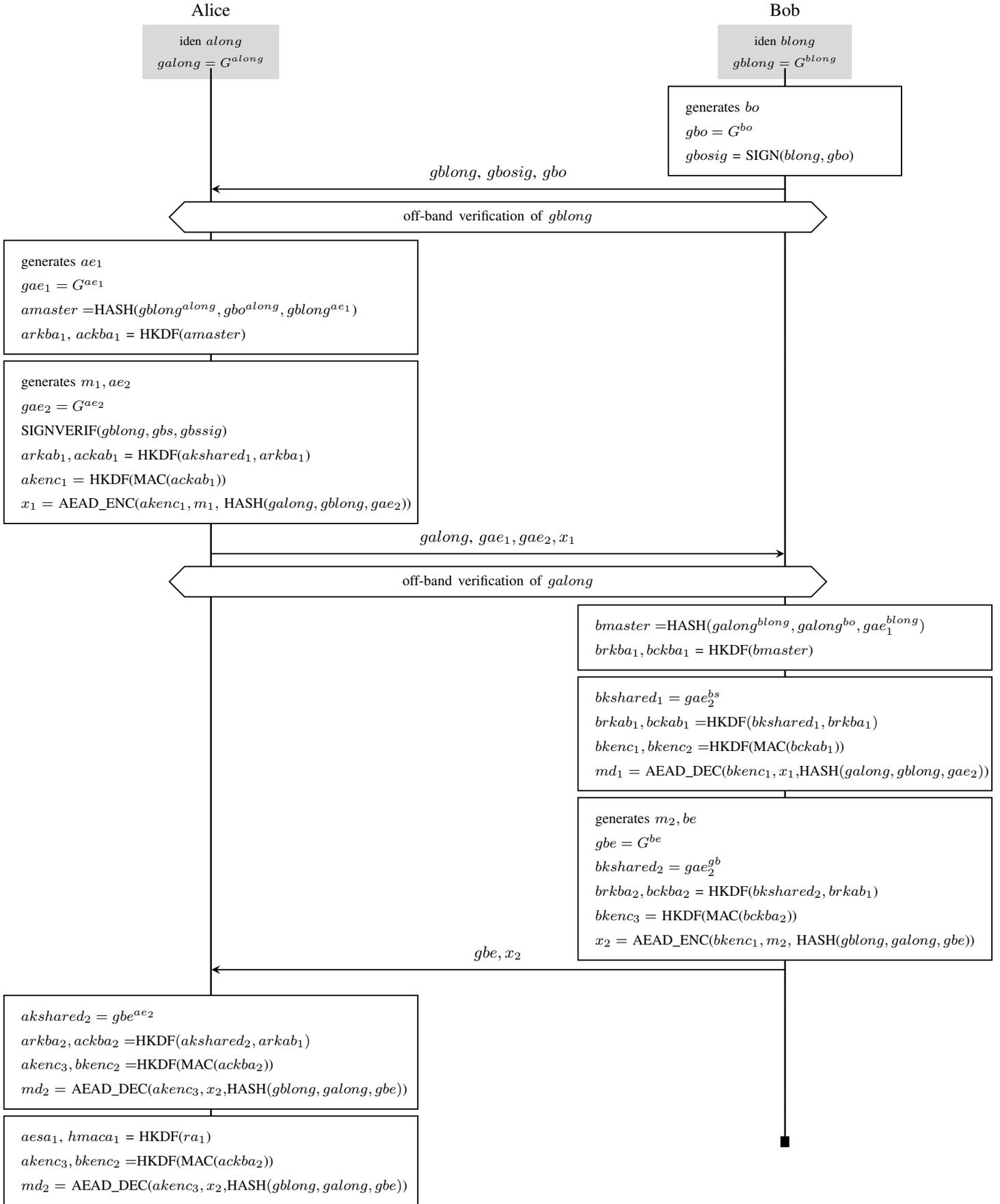

  \centering
  \footnotesize 
  \setmsckeyword{}
  \drawframe{no}    
  \begin{msc}[
    /msc/title top distance=0cm,
    /msc/first level height=.1cm,
    /msc/last level height=0.3cm,  
    /msc/head height=0cm,
    /msc/instance width=0cm,
    /msc/head top distance=0.5cm,
    /msc/foot distance=-0.0cm,
    /msc/instance width=0cm,
    /msc/condition height=0.1cm  
    ]{}
    \setlength{\instwidth}{0\mscunit} 
    \setlength{\instdist}{11cm}  

    \declinst{A}{              
      \begin{tabular}[c]{c}
        Alice \\ 
        \colorbox{gris}{\scriptsize{$\begin{array}{c} \text{iden }along \\ galong = G^{along}\end{array}$}} 
      \end{tabular}
    }{}
    \declinst{B}{              
      \begin{tabular}[c]{c}
        Bob \\ 
        \colorbox{gris}{\scriptsize{$\begin{array}{c} \text{iden } blong \\ gblong = G^{blong} \end{array}$}} 
      \end{tabular}
    }{} 

    \nextlevel[0.4]  
    \action*{\parbox{120pt}{\footnotesize{
          $\begin{array}[c]{l}
             \text{generates $bo$}\\
             \text{$gbo = G^{bo}$}\\
             \text{$gbosig$ = SIGN($blong, gbo$)}\\
           \end{array}$
         }}}{B}
    \nextlevel[3.5]  
    \mess{\messLabel{$gblong$, $gbosig$, $gbo$}}{B}{A}
    \nextlevel[0.4]  
    \condition{\messLabel{{\footnotesize off-band verification of $gblong$}}}{B,A}
    \nextlevel[1.2]  
    \action*{\parbox{220pt}{\footnotesize{
          $\begin{array}[c]{l}
             \text{generates $ae_1$}\\
             \text{$gae_1 = G^{ae_1}$}\\
             \text{$amaster = $HASH($gblong^{along}, gbo^{along}, gblong^{ae_1}$)}\\
             \text{$arkba_1$, $ackba_1$ = HKDF($amaster$)}\\
           \end{array}$
         }}}{A}
    \nextlevel[4.5]  
    \action*{\parbox{220pt}{\footnotesize{
          $\begin{array}[c]{l}
             \text{generates $m_1, ae_2$}\\
             \text{$gae_2 = G^{ae_2}$}\\
             \text{SIGNVERIF($gblong, gbs, gbssig$)}\\
             \text{$arkab_1, ackab_1$ = HKDF($akshared_1, arkba_1$)}\\
             \text{$akenc_1 = $ HKDF(MAC($ackab_1$))}\\
             \text{$x_1 = $ AEAD\_ENC($akenc_1,m_1,$ HASH($galong, gblong, gae_2$))}\\
           \end{array}$
         }}}{A}
    \nextlevel[7]  
    \mess{\messLabel{$galong$, $gae_1,gae_2,x_1$}}{A}{B}
    \nextlevel[0.4]  
    \condition{\messLabel{{\footnotesize off-band verification of $galong$}}}{B,A}
    \nextlevel[1.2]  
    \action*{\parbox{220pt}{\footnotesize{
       $\begin{array}[c]{l}
         \text{$bmaster = $HASH$(galong^{blong}, galong^{bo}, gae_1^{blong})$} \\
         \text{$brkba_1, bckba_1$ = HKDF($bmaster$)}
        \end{array}$
       }}}{B}
    \nextlevel[2.5]  
    \action*{\parbox{220pt}{\footnotesize{
       $\begin{array}[c]{l}
         \text{$bkshared_1 = gae_2^{bs}$} \\
         \text{$brkab_1, bckab_1 = $HKDF$(bkshared_1,brkba_1)$} \\
         \text{$bkenc_1, bkenc_2 = $HKDF(MAC($bckab_1$))} \\
         \text{$md_1 = $ AEAD\_DEC($bkenc_1, x_1, $HASH$(galong, gblong, gae_2)$)}
        \end{array}$
       }}}{B}
    \nextlevel[4.5]  
    \action*{\parbox{220pt}{\footnotesize{
          $\begin{array}[c]{l}
             \text{generates $m_2, be$}\\
             \text{$gbe = G^{be}$}\\
             \text{$bkshared_2 = gae_2^{gb}$}\\
             \text{$brkba_2, bckba_2$ = HKDF($bkshared_2, brkab_1$)}\\
             \text{$bkenc_3 = $ HKDF(MAC($bckba_2$))}\\
             \text{$x_2 = $ AEAD\_ENC($bkenc_1,m_2,$ HASH($gblong, galong, gbe$))}\\
           \end{array}$
         }}}{B}
    \nextlevel[6.5]  
    \mess{\messLabel{$gbe, x_2$}}{B}{A}
    \nextlevel]
    \action*{\parbox{220pt}{\footnotesize{
       $\begin{array}[c]{l}
         \text{$akshared_2 = gbe^{ae_2}$} \\
         \text{$arkba_2, ackba_2 = $HKDF$(akshared_2,arkab_1)$} \\
         \text{$akenc_3, bkenc_2 = $HKDF(MAC($ackba_2$))} \\
         \text{$md_2 = $ AEAD\_DEC($akenc_3, x_2, $HASH$(gblong, galong, gbe)$)}
        \end{array}$
       }}}{A}
    \nextlevel[4.5]  
    \action*{\parbox{220pt}{\footnotesize{
       $\begin{array}[c]{l}
         \text{$aesa_1$, $hmaca_1$ = HKDF($ra_1$)} \\
         \text{$akenc_3, bkenc_2 = $HKDF(MAC($ackba_2$))} \\
         \text{$md_2 = $ AEAD\_DEC($akenc_3, x_2, $HASH$(gblong, galong, gbe)$)}
        \end{array}$
       }}}{A}
   \end{msc}
   \vspace{-15pt}
  \caption{Complete Olm Protocol including optional pre-key signing, describing the initialization and transmission of two messages. Brackets around a message in transit indicates off-band verification, where an attacker can observe but not modify the message.}
  \label{fig:olm-handshake}
\end{figure*}

\begin{table*}[ht]
\centering
\begin{tabular}{|l|l|l|}
\hline
\textbf{Protocol} & \textbf{Olm} \cite{olm-docs} & \textbf{Signal} \cite{fm-signal} \\ \hline
Identity Key    & Curve25519   & X25519   \\
Fingerprint Key & Ed25519 & N/A           \\
Pre-keys        & Curve25519   & X25519    \\
Ephemeral Key & N/A & X25519          \\
Signed Pre-keys  & (Optional) Curve25519     & X25519         \\
Key Exchange   & Triple Diffie-Hellman (3DH) & Extended Triple Diffie-Hellman (X3DH) \\
Ratcheting & Double Ratchet & Double Ratchet \\ \hline
\end{tabular}
\caption{Comparison of Olm and Signal} \label{tab:p2p_comparison}
\end{table*}

\textbf{Megolm}. Megolm is used to ensure group communications are encrypted. Originally introduced in 2016, before Sender Key (group messaging variant of Signal) \cite{sender-keys} was published, Megolm has very similar functionality to \textit{Sender Keys}.  However, Megolm utilizes Olm as opposed to Signal as the secure peer-to-peer channel in order to share Megolm session information between peers. Additionally, Megolm utilizes a custom ratchet scheme that can be advanced an arbitrary amount forward while needing at most 1020 hash computations \cite{megolm-docs}, which is further described in Figure \ref{sub:Megolm Ratchet Definition}. Aside from the details in cryptographic primitives between implementations of Sender Keys and Megolm as shown in Table \ref{tab:sk_comparison}, the secure channel protocol, and the ratcheting scheme, all other details between the two protocols remain the same.


\begin{table*}[h!]
\centering
\begin{tabular}{|>{\centering\arraybackslash}m{3.0cm}|>{\centering\arraybackslash}m{2cm}|>{\centering\arraybackslash}m{2cm}|>{\centering\arraybackslash}m{2cm}|>{\centering\arraybackslash}m{2cm}|}
\hline
\textbf{Protocol} & \textbf{P2P Encryption Protocol} & \textbf{Message Encryption} & \textbf{Ratchet Mechanism} & \textbf{Signature Key} \\ 
\hline
Megolm \cite{megolm-docs} & Olm & AES-256 (CBC) & Custom & ED25519 \\ 
WhatsApp \cite{wb-whatsapp}  & Signal & AES-256 (CBC) & HMAC-SHA-256 & Curve25519 \\ 
Signal Sender Keys \cite{signal_app_sk} & Signal & AES-256 (CBC) & HMAC-SHA-256 & Curve25519 \\ 
Session Application \cite{session_blog} & Signal & AES-256 (CBC) & HMAC-SHA-256 & ED25519 \\ 
Facebook Messenger \cite{FBMessenger_2023} & Signal & AES-256 (CBC) & HMAC-SHA-256 & Curve25519 \\ 
\hline
\end{tabular}
\caption{Comparison of Sender Keys implementations and Megolm}
\label{tab:sk_comparison}
\end{table*}

For precise game-theoretic cryptographic definitions of Olm and Megolm, we defer to \cite{Albrecht2024}, of which our models and analysis are closely based upon. Our derived Olm and Megolm handshakes are shown in Figures \ref{fig:olm-handshake} and \ref{fig:megolm-sk}, respectively.

\subsection{Security of Matrix Cryptographic Suite}%
\label{sec:bk_sec}
Since Olm and Megolm have similar security goals as Signal and Sender Keys, we discuss the security of all four protocols.

\textbf{Signal}. The Signal protocol has seen ample formal analysis since its inception in 2013, including both symbolic \cite{vp} \cite{nadim-automated} \cite{Bhargavan_dy} \cite{Bhargavan_Jacomme_Kiefer_Schmidt} and computational \cite{Alwen_Coretti_Dodis_2019} \cite{fm-signal} analysis. These analyses demonstrated the post-compromise and forward secrecy guarantees of the double ratchet scheme and the authentication and security guarantees of the X3DH handshake in the manual, computational setting \cite{fm-signal} and the symbolic, mechanized setting \cite{nadim-automated} \cite{fm-signal-cm}.

\textbf{Sender Keys}. Sender Keys, in comparison with Signal, is relatively understudied, as there only exists a single computational analysis \cite{Balbas_Collins_Gajland_2023} and, to the best of our knowledge, no mechanical nor symbolic analysis. Balbas et al. formally define a "Group Messenger" primitive and the Sender Keys protocol, then they prove the security of the Sender Keys protocol with respect to their custom primitive. In the post-compromise security case, Balbas et al. observe leaking the sender key can result an outside attacker being able to forge a message. To prevent this, the authors propose \textit{Sender Keys+} which, among other non-cryptographic improvements, proposes MACing message contents with a hash of the chain key (which prevents leaked message keys from exposing the corresponding messages), and ratcheting signature keys in addition to ratchet keys (which prevents unauthorized injections in the case the identity key is compromised). The authors prove the security of \textit{Sender Keys+} by hand.

\textbf{Olm and Megolm}. There has been just a single \textit{computational} formal analysis of Olm and Megolm \cite{Albrecht2024} and, to the best of our knowledge, no mechanical nor symbolic analysis. Albrecht et al. formally define a custom primitive, \textit{Device-Oriented Group Messaging (DOGM)}, and prove the security and authentication of the composition of Olm and Megolm with respect to it by hand. The authors also describe some trivial attacker wins with respect to their threat model; compromising the long-term Megolm session compromises the messages, compromising a Megolm identity key allows the initialization of forged Olm sessions, compromising a Megolm identity allows the adversary to forge messages, and compromising a Megolm identity allows an adversary to initiate malicious epochs.

\subsection{Scope of Our Analysis}%
\label{sec:Scope}

In our analysis, we take a \textit{symbolic} approach to reasoning about Olm and Megolm. Our analysis is in \textit{complement} to the previous formal, computational analysis of Olm and Megolm \cite{Albrecht2024}. 
We also seek to \textit{mechanize} the verification of Olm and Megolm using a computer-aided cryptographic analysis tool, complementing the manual on-paper proofs provided in \cite{Albrecht2024}. Using mechanized proofs and computer-aided analysis is particularly necessary for cryptographic protocols due to not just their importance, but the high level of rigor and precision required to correctly specify them \cite{Liao_sok}. Hand-written proofs are error-prone and often carry subtle mistakes that go unnoticed. This has consistently shown to be the case for both mathematical literature \cite{proofs-generations} \cite{Lamport_2012} and distributed protocol literature \cite{Zave_2012}. Thus, we stress the importance of mechanization for our analysis.

\definecolor{gris}{gray}{0.85}
\begin{figure*}[ht!]
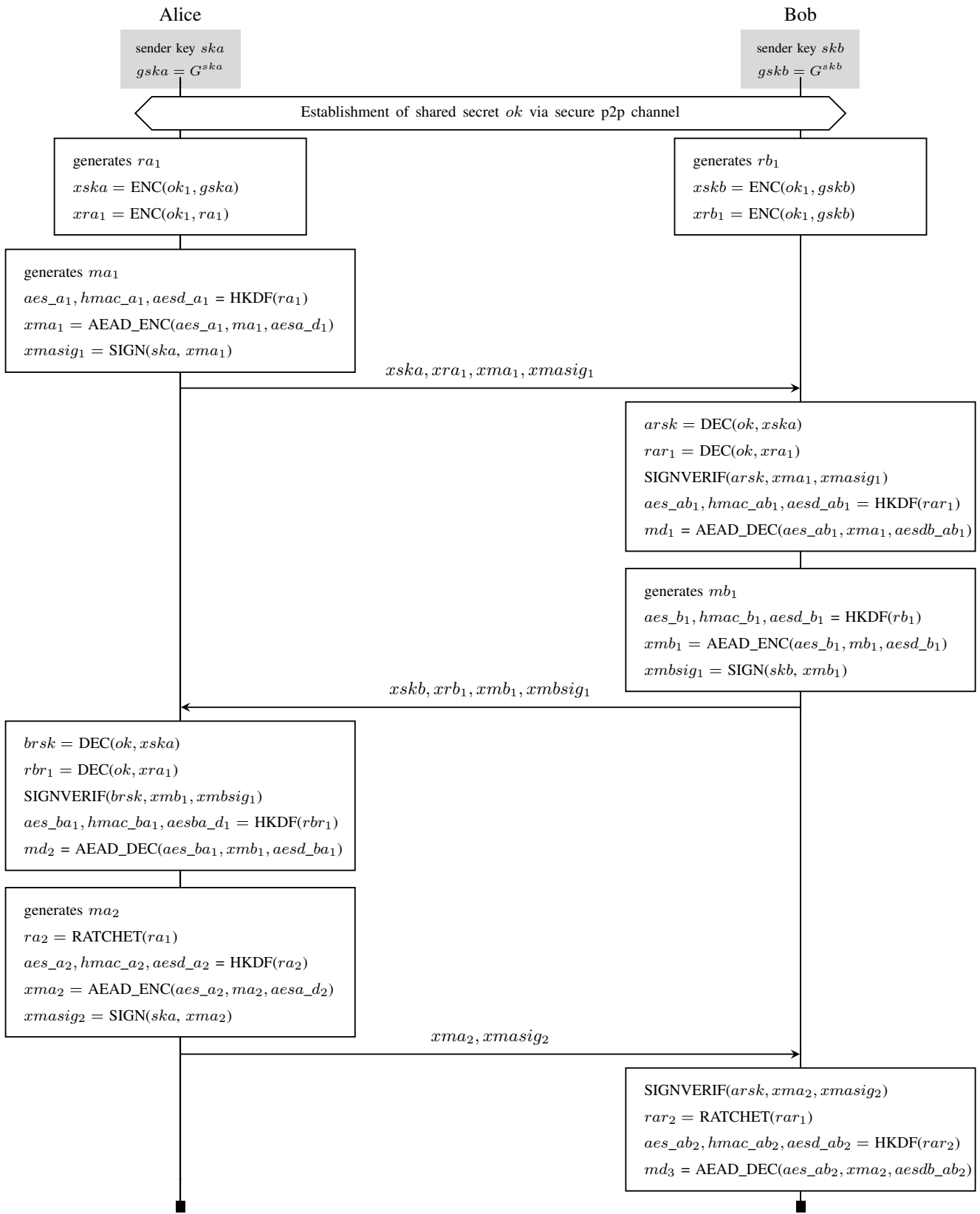

  \centering
  \footnotesize 
  \setmsckeyword{}
  \drawframe{no}    
  \begin{msc}[
    /msc/title top distance=0cm,
    /msc/first level height=.1cm,
    /msc/last level height=0.3cm,  
    /msc/head height=0cm,
    /msc/instance width=0cm,
    /msc/head top distance=0.0cm,
    /msc/foot distance=-0.0cm,
    /msc/instance width=0cm,
    /msc/condition height=0.1cm  
    ]{}
    \setlength{\instwidth}{0\mscunit} 
    \setlength{\instdist}{11cm}  

    \declinst{A}{              
      \begin{tabular}[c]{c}
        Alice \\
        \colorbox{gris}{\scriptsize{$\begin{array}{c} \text{sender key }ska \\ gska = G^{ska} \end{array}$}} 
      \end{tabular}
    }{}
    \declinst{B}{              
      \begin{tabular}[c]{c}
        Bob \\
        \colorbox{gris}{\scriptsize{$\begin{array}{c} \text{sender key } skb \\ gskb = G^{skb} \end{array}$}} 
      \end{tabular}
    }{} 

    \nextlevel[0.4]  
    \condition{\messLabel{{\footnotesize Establishment of shared secret $ok$ via secure p2p channel}}}{B,A}
    \nextlevel[1.2]  
    \action*{\parbox{120pt}{\footnotesize{
          $\begin{array}[c]{l}
             \text{generates $ra_1$}\\
             \text{$xska = $ ENC($ok_1,gska$)} \\
             \text{$xra_1 = $ ENC($ok_1,ra_1$)}
           \end{array}$
         }}}{A}
    \action*{\parbox{120pt}{\footnotesize{
          $\begin{array}[c]{l}
             \text{generates $rb_1$}\\
             \text{$xskb = $ ENC($ok_1,gskb$)} \\
             \text{$xrb_1 = $ ENC($ok_1,gskb$)}
           \end{array}$
    }}}{B}
    \nextlevel[3.5]  
    \action*{\parbox{170pt}{\footnotesize{
          $\begin{array}[c]{l}
             \text{generates $ma_1$}\\
             \text{$aes\_a_1,hmac\_a_1,aesd\_a_1$ = HKDF($ra_1$)} \\
             \text{$xma_1 = $ AEAD\_ENC($aes\_a_1, ma_1, aesa\_d_1$)} \\
             \text{$xmasig_1 = $ SIGN($ska$, $xma_1$)}
           \end{array}$
         }}}{A}
    \nextlevel[4.75]  
    \mess{\messLabel{$xska, xra_1, xma_1,xmasig_1$}}{A}{B}
    \nextlevel[0.4]  
    \action*{\parbox{170pt}{\footnotesize{
          $\begin{array}[c]{l}
             \text{$arsk = $ DEC($ok, xska$)}\\
             \text{$rar_1 = $ DEC($ok, xra_1$)}\\
             \text{SIGNVERIF($arsk, xma_1, xmasig_1$)} \\
             \text{$aes\_ab_1, hmac\_ab_1, aesd\_ab_1 = $ HKDF($rar_1$)}\\
             \text{$md_1$ = AEAD\_DEC($aes\_ab_1, xma_1, aesdb\_ab_1$)}
           \end{array}$
         }}}{B}
    \nextlevel[5.5]  
    \action*{\parbox{170pt}{\footnotesize{
          $\begin{array}[c]{l}
             \text{generates $mb_1$}\\
             \text{$aes\_b_1,hmac\_b_1,aesd\_b_1$ = HKDF($rb_1$)} \\
             \text{$xmb_1 = $ AEAD\_ENC($aes\_b_1, mb_1, aesd\_b_1$)} \\
             \text{$xmbsig_1 = $ SIGN($skb$, $xmb_1$)}
           \end{array}$
         }}}{B}
    \nextlevel[4.75]  
    \mess{\messLabel{$xskb, xrb_1, xmb_1, xmbsig_1$}}{B}{A}
    \nextlevel[0.4]  
    \action*{\parbox{170pt}{\footnotesize{
          $\begin{array}[c]{l}
             \text{$brsk = $ DEC($ok, xska$)}\\
             \text{$rbr_1 = $ DEC($ok, xra_1$)}\\
             \text{SIGNVERIF($brsk, xmb_1, xmbsig_1$)} \\
             \text{$aes\_ba_1, hmac\_ba_1, aesba\_d_1 = $ HKDF($rbr_1$)}\\
             \text{$md_2$ = AEAD\_DEC($aes\_ba_1, xmb_1, aesd\_ba_1$)}
           \end{array}$
         }}}{A}
    \nextlevel[5.5]  

   \end{msc}
   \vspace{5pt}
  \caption{Complete Megolm/Sender Keys protocol, communicating 3 messages between Alice and Bob. Note, the constructions of Megolm and Sender Keys are the same besides (1) different ratcheting algorithms, and (2) different secure P2P channel protocols.}
  \label{fig:megolm-sk}
\end{figure*}

\section{Olm and Megolm Formal Models}%
\label{sec:Our Model}

We fully model the Matrix protocol suite as well as Sender Keys using \pv, an automated symbolic analysis tool in the Dolev-Yao model.  
In this section we overview the \pv automated symbolic reasoning tool, the subsequent abstractions and limitations of our models, the details of our models, and the properties of interest we seek to reason about.





\subsection{Modeling in ProVerif}%
\label{sub:Modeling in ProVerif}
We choose \pv as our symbolic protocol analysis tool of choice, boasting an efficient pi-calculus-based proof engine which reduces the term rewriting search problem that characterizes symbolic cryptanalysis to horn clauses, as well as an expressive syntax for specifying protocols \cite{Blanchet_Smyth_Cheval_Sylvestre}. 

To model protocols in \pv, we begin by defining each participant as a process, representing the roles they play within the protocol. Each process consists of message exchanges, including cryptographic operations such as encryption, decryption, signing, and verification. Cryptographic primitives are modeled symbolically using predefined functions in \pv, which capture their idealized behavior without considering their computational implementation.

The adversary is explicitly modeled with the ability to intercept, modify, and inject messages into the communication channel. However, they are restricted by the symbolic security assumptions: they cannot invert cryptographic functions unless explicitly allowed by the model. These symbolic constraints allow us to analyze potential attacks, such as replay attacks, man-in-the-middle attacks, or key compromise impersonation, within the Dolev-Yao framework.

In addition to the processes and messages, we specify the security properties of interest using queries, which \pv attempts to verify or refute. Typical properties include authentication, secrecy, and integrity. These properties are formally expressed in terms of reachability, correspondence assertions, or observational equivalence, depending on the specific security goals of the protocol under analysis.

By following this structured modeling methodology, we ensure that the symbolic model accurately reflects the intended behavior of the protocol, while accounting for the adversarial capabilities. This allows for the comprehensive verification of the security properties under consideration. Additionally, in order to ensure \pv is always able to complete a proof (or discover a counterexample), we ran our experiments on a research computing cluster with 88 cascade lake CPUs and 5TB of RAM.

\subsection{Model Details}%
\label{sub:Model Details}

To construct our \pv models, we carefully and exhaustively analyzed the existing specifications of Olm \cite{olm-docs}, Megolm \cite{megolm-docs}, and Sender Keys \cite{FBMessenger_2023} \cite{wb-whatsapp}, the canonical implementations of Olm and Megolm provided by the Matrix foundation \cite{vodozemac} \cite{libolm}, the Matrix protocol specification \cite{matrix-spec}, and much of the relevant online discussion \cite{soatok2024gist}. We also closely compared our symbolic models with the existing computational constructions of Olm, Megolm, and Sender Keys \cite{Albrecht2024} \cite{Balbas_Collins_Gajland_2023} to best ensure our models accurately capture the intended behaviors of the protocols. Additionally, although not central to our analysis, we construct \vp \cite{vp} models of of Olm, Megolm, and Sender Keys for the general benefit of the community.

\textbf{Olm}. We model the complete cryptographic handshake and the ratcheting mechanism of Olm. A diagram of the modeled handshake is shown in Appendix Section \ref{fig:olm-handshake}.

\textbf{Megolm}. We model the initialization of a Megolm session, the transmission of the ratchet key to peers in the Megolm session via Olm, the exchange of messages between peers in a Megolm session, and the incrementation of the Megolm ratchet. A diagram of the modeled cryptographic exchange is shown in Figure \ref{fig:megolm-sk}.

\textbf{Signal}. We employ a previously created Signal model, as presented in \cite{nadim-automated} and visualized in .

\textbf{Sender Keys}. We model the initialization of a Sender Keys session, the transmission of the ratchet key to peers in the Sender Keys session via Signal, the exchange of messages between peers in a Sender Keys session, and the incrementation of the Sender Keys ratchet. A diagram of the modeled cryptographic exchange is shown in Figure \ref{fig:megolm-sk}.

\subsection{Abstractions and Limitations}%
\label{sub:Abstractions and Limitations}
Our models are fully faithful to the Olm, Megolm, and Sender Keys specifications, modulo the following abstractions and limitations.

\noindent~$\bullet$~\textbf{Perfect Underlying Primitives}. As in the symbolic cryptanalysis model, we assume the underlying cryptographic primitives are \textit{black-boxes}. That is, we assume the primitives used in our models are secure and correct, particularly \textsc{HKDF}, \textsc{HMAC}, \textsc{AEAD}, and key signing. \pv and other cryptographic reasoning in tools in the symbolic model traditionally assume perfect computational cryptography. \\
\noindent~$\bullet$~\textbf{Abstract Message Terms}. Our models are abstract with respect to message formats, treating each message as a symbolic constructor. This abstract treatment of protocol messages is typical of symbolic tools. \\
\noindent~$\bullet$~\textbf{No Ratchet Reseeding}. The Megolm ratchet is designed to re-seed after every $2^{24}$ iterations as described in the Appendix \ref{sub:Megolm Ratchet Definition}. We do not capture this reseeding in our model --- only forward iterations --- as it is not relevant to the analysis under the Dolev-Yao model. \\
\noindent~$\bullet$~\textbf{Finite Peers}. We reason about a finite number of peers of a Megolm or Sender Keys group in our model, and prove our properties of interest under that assumption. \\
\noindent~$\bullet$~\textbf{Perfect Message Ordering}. We do not reason about message orderings in Megolm nor Sender Keys, and assume all messages arrive in proper order.


\subsection{Properties of Interest}%
\label{sub:Properties of Interest}
Motivated by the Olm, Signal, Megolm, and Sender Keys specifications and previous literature, we are concerned with the following properties:

\noindent~$\bullet$~\textbf{Olm Message Confidentiality and Authentication}. All messages transmitted by Olm, assuming pre-keys are signed, should be confidential and authenticated, given an attacker with full control over the communication channel. This property is explicitly derived from the Olm specification \cite{olm-docs} and was additionally proven with pen and paper in \cite{Albrecht2024}.

\noindent~$\bullet$~\textbf{Olm Session Key Confidentiality}. Each Olm session key remains confidential, given an attacker with full control over the communication channel. This property was explicitly derived from the Olm specification \cite{olm-docs}. 

\noindent~$\bullet$~\textbf{Olm Forward Secrecy}. Upon leaking Olm identity keys and Olm session keys and \textit{assuming} Olm chooses to sign its pre-key, messages from previous sessions should remain secure given an attacker with full control over the communication channel. This property is explicitly derived from the Olm specification \cite{olm-docs} and was additionally proven with pen and paper in \cite{Albrecht2024}.

\noindent~$\bullet$~\textbf{Olm Post-Compromise Security}. Upon leaking Olm identity keys, all future messages should remain secure, given an attacker with full control over the communication channel. This property is explicitly derived from the Olm specification \cite{olm-docs}. 

\noindent~$\bullet$~\textbf{Megolm Message Confidentiality and Authentication}. All messages transmitted by Megolm should be confidential and authenticated, given an attacker with full control over the communication channel. This property was explicitly derived from the Megolm specification \cite{megolm-docs} and was additionally proven with pen and paper in \cite{Albrecht2024}.

\noindent~$\bullet$~\textbf{Megolm Forward Secrecy}. Upon leaking the ratchet key , all previous messages encrypted by symmetric keys derived from previous ratchets should remain secure. This property was explicitly derived from the Megolm specification \cite{megolm-docs}.

\section{Analysis}%
\label{sec:Results}

In this section we present our mechanized results, including the proofs of the properties of interest mentioned in the previous section \ref{sub:Properties of Interest}, various proofs and attacks from the protocol specifications and previous literature, and the results of our comparative analysis between Megolm
and Sender Keys.

\subsection{Proven Properties of Interest}%
\label{sub:Proofs}

Using our \pv models, we automatically prove the Olm and Megolm properties as stated and motivated in section ~\ref{sub:Properties of Interest}. This includes Olm message confidentiality and authentication, Olm session key confidentiality, Olm forward secrecy (assuming Olm chooses to sign pre-keys), Olm post-compromise security, Megolm message confidentiality and authentication, and Megolm forward secrecy.

\subsection{Mechanization of Manual Proofs and Attacks}%
\label{sub:Literature Mechanization}

Using our \vp models, we mechanize various proofs, attacks, and limitations as described in the literature and documentation. This allows us to extract precise, verifiable attack sequences, which we narrate over the rest of this section.

\subsubsection{Compromising a Megolm session compromises all transmitted messages via the Megolm session}
\label{sub:simple-megolm-comp}
As shown on pen and paper in the computational formulation of Matrix as a trivial attacker win, compromising a Megolm session compromises the ratchet key contained within it, allowing an attacker to decrypt all messages sent with the Megolm session after the compromise \cite{Albrecht2024}.

\subsubsection{Unknown Key-share in Olm}%
\label{sub:Unknown Key-share in Olm}

Olm, like Signal, is reliant on the off-band verification of the pairwise identity keys. If this verification does not occur, Olm is vulnerable to an unknown key-share attack, as described in the NCC audit of Olm and Megolm \cite{ncc-audit}. This attack works as follows. Alice initiates an Olm session with Bob. Bob tries to send his identity and his (signed or unsigned) pre-key to Alice (through the server), but Mallory replaces Bob's identity key and pre-key, signed or unsigned accordingly, with her own. Alice computes the Olm pairwise master key with the malicious values, uses the master key to derive a ratchet key, derives a symmetric key from the ratchet key, and encrypts a message. Alice then transmits her identity key and pre-key to Bob, which are observed by Mallory. Then, using Alice's identity and pre-key, Mallory computes the pairwise, master key, derives the ratchet key, derives the message key, and decrypts the message.

\subsubsection{Unknown Key-share in Megolm}%
\label{sub:subsection name}

At a high level, this attack involves the \textit{propagation} of the Olm unknown key-share attack \textit{through} Megolm. That is, an unknown key-share attack in an Olm session will compromise the Megolm session. This attack works as follows. Alice and Bob intend to initialize a Megolm connection. Alice initializes a Olm session with Bob; however, the Olm session, particularly the pairwise master key, is compromised via an unknown key-share attack as described in \ref{sub:Unknown Key-share in Olm}. Alice then generates her Megolm session information, including her sender key pair and her ratchet key. Using the compromised Olm master key, Alice derives a ratchet key, and a message key from the ratchet key. Alice encrypts her Megolm session information using the message key and sends it to Bob. Mallory captures the encrypted Alice's Megolm session, derives the Olm message key from the Olm master key she captured, and decrypts Alice's Megolm session. Now, with access to the Megolm ratchet key in Alice's Megolm session, Mallory can decrypt all the messages Alice sends to Bob via the Megolm session.

\subsubsection{Compromising an Olm identity key without pre-key signing compromises the messages of an Olm exchange before the first ratchet key advancement}%
\label{sub:Olm PCS without pre-key signing}
As stated in the Olm specification, choosing whether to sign pre-keys or not is left to the protocol implementer \cite{olm-docs}. The Olm authors cite a tradeoff between forward secrecy and deniability \cite{olm-docs}. As shown in the computational formulation of Megolm \cite{Albrecht2024}, we verify leaving pre-keys unsigned does indeed compromise forward secrecy. The attack works as follows. Alice initiates an Olm session with Bob. Bob tries to send his identity key and his unsigned pre-key to Alice (through the server), but Mallory replaces Bob's pre-key with her own. Alice, using her identity key and pre-key, computes the pairwise Olm session key with Bob's identity key and Mallory's pre-key. Alice uses the Olm pairwise master key to derive a ratchet key, derive a symmetric key from the ratchet key, then encrypts a message with it. Alice transmits her identity key and pre-key to Bob, which Mallory captures but does not modify in transit. Then, Bob's private key gets leaked to Mallory. Mallory then derives the pairwise Olm master key by performing Diffie-Hellman with (1) Alice's public key and Bob's private key, (2) Alice's public key and Mallory's injected pre-key, and (3) Alice's transmitted public pre-key and Bob's private key. Using the pairwise Olm master key, Mallory derives the message key and decrypts the message Alice sent to Bob. We do observe, however, that the Olm session \textit{heals} after Alice and Bob advance their ratchet key, exchanging keys for Diffie-Hellman to do so. Thus, this attack only compromises the first (or first few) messages. A visualization of this attacker sequence is given in Figure \ref{fig:olm-pcs-vis}.

\definecolor{gris}{gray}{0.85}
\begin{figure*}[ht!]
  \centering
  \footnotesize 
  \setmsckeyword{}
  \drawframe{no}    
  \begin{msc}[
    /msc/title top distance=0cm,
    /msc/first level height=.1cm,
    /msc/last level height=0.3cm,
    /msc/head height=0cm,
    /msc/instance width=0cm,
    /msc/head top distance=0.5cm,
    /msc/foot distance=-0.0cm,
    /msc/instance width=0cm,
    /msc/condition height=0.1cm 
    ]{}
    \setlength{\instwidth}{0\mscunit} 
    \setlength{\instdist}{5cm}  

    \declinst{A}{              
      \begin{tabular}[c]{c}
        Alice \\ 
        \colorbox{gris}{\scriptsize{$\begin{array}{c} \text{iden }along \\ galong = G^{along}\end{array}$}} 
      \end{tabular}
    }{}
    \declinst{M}{              
      \begin{tabular}[c]{c}
        Attacker \\ 
      \end{tabular}
    }{}
    \declinst{B}{              
      \begin{tabular}[c]{c}
        Bob \\ 
        \colorbox{gris}{\scriptsize{$\begin{array}{c} \text{iden } blong \\ gblong = G^{blong} \end{array}$}} 
      \end{tabular}
    }{} 
    \nextlevel[0.75]  
    \action*{\parbox{120pt}{\footnotesize{
          $\begin{array}[c]{l}
             \text{generates $bo$}\\
             \text{$gbo = G^{bo}$}\\
           \end{array}$
         }}}{B}
    \nextlevel[3.5]  
    \mess{\messLabel{$gblong$, $gbo$}}{B}{M}
    \nextlevel[0.4]  
    \action*{\parbox{120pt}{\footnotesize{
          $\begin{array}[c]{l}
             \text{replaces $gbo$ with $gbm$}\\
           \end{array}$
         }}}{M}
    \nextlevel[2.5]  
    \mess{\messLabel{$gblong$, $gbm$}}{M}{A}
    \nextlevel[0.3]  
    \condition{\messLabel{{\footnotesize off-band verification of $gblong$}}}{B,M,A}
    \nextlevel[1.4]  
    \action*{\parbox{220pt}{\footnotesize{
          $\begin{array}[c]{l}
             \text{generates $ae_1$}\\
             \text{$gae_1 = G^{ae_1}$}\\
             \text{$amaster = $HASH($gblong^{along}, gbm^{along}, gblong^{ae_1}$)}\\
             \text{$arkba_1$, $ackba_1$ = HKDF($amaster$)}\\
           \end{array}$
         }}}{A}

    \nextlevel[4.75]  
    \action*{\parbox{220pt}{\footnotesize{
          $\begin{array}[c]{l}
             \text{generates $m_1, ae_2$}\\
             \text{$gae_2 = G^{ae_2}$}\\
             \text{$arkab_1, ackab_1$ = HKDF($akshared_1, arkba_1$)}\\
             \text{$akenc_1 = $ HKDF(MAC($ackab_1$))}\\
             \text{$x_1 = $ AEAD\_ENC($akenc_1,m_1,$ HASH($galong, gblong, gae_2$))}\\
           \end{array}$
         }}}{A}
    \nextlevel[6.75]  
    \mess{\messLabel{$galong$, $gae_1,gae_2,x_1$}}{A}{M}
    \mess{\messLabel{$galong$, $gae_1,gae_2,x_1$}}{M}{B}
    \nextlevel[0.4]  
    \condition{\messLabel{{\footnotesize off-band verification of $galong$}}}{A,M,B}
    \nextlevel[1.4]  
    \action*{\parbox{220pt}{\footnotesize{
       $\begin{array}[c]{l}
         \text{$bmaster = $HASH$(galong^{blong}, galong^{bo}, gae_1^{blong})$} \\
         \text{$brkba_1, bckba_1$ = HKDF($bmaster$)}
        \end{array}$
       }}}{B}
    \nextlevel[2.75]  
    \action*{\parbox{220pt}{\footnotesize{
       $\begin{array}[c]{l}
         \text{$bkshared_1 = gae_2^{bs}$} \\
         \text{$brkab_1, bckab_1 = $HKDF$(bkshared_1,brkba_1)$} \\
         \text{$bkenc_1, bkenc_2 = $HKDF(MAC($bckab_1$))} \\
         \text{$md_1 = $ AEAD\_DEC($bkenc_1, x_1, $HASH$(galong, gblong, gae_2)$)}
        \end{array}$
       }}}{B}
    \nextlevel[5.5]  
    \mess{\messLabel{\textbf{Leaks} $blong$}}{B}{M}
    \nextlevel[0.75]  
    \action*{\parbox{230pt}{\footnotesize{
          $\begin{array}[c]{l}
             \text{$bmaster$ = HASH($blong^{galong}$, $gblong^{gbm}$, $gae_1^{blong}$)}\\
             \text{$brkba_1$, $bckba_1$ = HKDF($bmaster$)}\\
             \text{$bcmba_1$ = HKDF(MAC($bckba_1$))}\\
             \text{$m_1 = $ AEAD\_DEC($x_1$, $bcmba_1$, HASH($galong, gblong, gae_2$))} \\
             \text{\textbf{Obtains} $m_1$}
           \end{array}$
         }}}{M}


   \end{msc}
   \vspace{-5pt}
   \caption{Post-compromise attacker sequence on unsigned Olm key exchange, as described in \ref{sub:Olm PCS without pre-key signing}}
  \label{fig:olm-pcs-vis}
\end{figure*}

\subsubsection{Compromising an Olm identity key with pre-key signing preserves post-compromise security}%
\label{sub:Olm PCS with pre-key signing}
We prove this property with respect to an active and passive attacker. Intuitively, pre-key signing prevents an attacker from replacing the pre-keys with malicious ones; thus, the future compromise of an identity key will not compromise the messages of the associated Olm session. Again note, this is the default behavior of the Signal protocol, as was proven in \cite{vp} \cite{fm-signal} \cite{fm-signal-cm}.

\subsubsection{Megolm \& Sender Keys message injection after sender key compromise}%
\label{sub:sk-auth-fs}
As pointed out in the computational formulation of Sender Keys, compromising a sender key pair allows the attacker to inject messages that correspond to key material from the respective Megolm session that were used before the sender key was compromised \cite{Balbas_Collins_Gajland_2023}. We verify this is indeed the case. The attack works as follows. Alice securely initiates an Olm session with Bob. Alice generates her Megolm session containing her ratchet key and public sender key, then sends it to Bob via Olm. Alice sends a few messages to Bob, advancing the ratchet key as such. Alice's sender key is then compromised by Mallory. At this point, Mallory can inject a message corresponding to previous key material used before the sender key was compromised, signing the message with the current sender key. Note, Megolm peers will store a copy of the earliest known ratchet value \cite{megolm-docs}, thus this injection is feasible. 

\subsubsection{The Sender Keys extension proposed by Balbas et al. \cite{Balbas_Collins_Gajland_2023} maintains authenticity and confidentiality} Among other control-related changes not relevant to analysis under Dolev-Yao, the Sender Keys extension as proposed by Balbas et al. proposes ratcheting the sender key with a hash algorithm in order to maintain authentication forward secrecy and using MAC as an alternative to AEAD when computing the chain key \cite{Balbas_Collins_Gajland_2023}. They prove their extension satisfies authentication and confidentiality. We replicate this proof with respect to an active and passive attacker.

\subsection{Megolm vs. Sender Keys Post-Compromise Security}%
\label{sub:Comparison}
Most notably in our analysis, we are the first to observe and mechanically prove that the post-compromise security of Megolm is explicitly weaker than in Sender Keys due to the differences in the forward secrecy guarantees of the respective peer-to-peer secure channels. However, this is the only major difference in symbolic security we discover in our analysis.

\subsubsection{Olm identity key compromise without pre-key signing compromises the reliant Megolm session}%
\label{sub:OlmMegolm PCS attack}
Critically, we observe that Olm \textit{only} transmits Megolm session data \cite{olm-docs}. So, the compromise of \textit{any} Olm messages will compromise Megolm session data. Similarly to how the unknown key-share attack on Megolm is caused by the unknown key-share attack in Olm, the leakage of Olm messages via the attack described in \ref{sub:Olm PCS without pre-key signing} is sufficient to leak Megolm session data. 

The attack works as follows. Alice and Bob intend to initialize a Megolm connection. Alice initiates an Olm session with Bob. Bob tries to send his identity key and his unsigned pre-key to Alice (through the server), but Mallory replaces Bob's pre-key with her own. Alice, using her identity key and pre-key, computes the pairwise Olm session key with Bob's identity key and Mallory's pre-key. Alice uses the Olm pairwise master key to derive a ratchet key, derive a symmetric key from the ratchet key, then encrypts a message with it. Alice transmits her identity key and pre-key to Bob, which Mallory captures but does not modify in transit. Alice then generates her Megolm session information, including her sender key pair and her ratchet key. Using the Olm pairwise master key, Alice derives an Olm ratchet key, and an Olm message key from the ratchet key. Alice encrypts her Megolm session information using the Olm message key and sends it to Bob. Mallory captures the encrypted Megolm session. Alice sends Megolm messages to Bob via her Megolm session as normal. Then, Bob's private key gets leaked to Mallory. Mallory then derives the pairwise Olm master key from doing Diffie-Hellman with (1) Alice's public key and Bob's private key, (2) Alice's public key and Mallory's injected pre-key, and (3) Alice's transmitted public pre-key and Bob's private key. Using the Olm pairwise master key, Mallory derives the Olm message key and decrypts the Megolm session message Alice sent to Bob. Using the decrypted Megolm session, Mallory can now read the previous messages Alice sent to Bob (as Mallory recorded them), and all the future Alice sends to Bob via the Megolm session. A visualization of this attacker sequence is given in Figure \ref{fig:megolm-pcs-vis}.

\definecolor{gris}{gray}{0.85}
\begin{figure*}[ht!]
  \centering
  \footnotesize 
  \setmsckeyword{}
  \drawframe{no}    
  \begin{msc}[
    /msc/title top distance=0cm,
    /msc/first level height=.1cm,
    /msc/last level height=0.3cm,  
    /msc/head height=0cm,
    /msc/instance width=0cm,
    /msc/head top distance=0.5cm,
    /msc/foot distance=-0.0cm,
    /msc/instance width=0cm,
    /msc/condition height=0.1cm  
    ]{}
    \setlength{\instwidth}{0\mscunit} 
    \setlength{\instdist}{5cm}  

    \declinst{A}{              
      \begin{tabular}[c]{c}
        Alice \\ 
        \colorbox{gris}{\scriptsize{$\begin{array}{c} \text{iden }along \\ galong = G^{along}\end{array}$}} 
      \end{tabular}
    }{}
    \declinst{M}{              
      \begin{tabular}[c]{c}
        Attacker \\ 
      \end{tabular}
    }{}
    \declinst{B}{              
      \begin{tabular}[c]{c}
        Bob \\ 
        \colorbox{gris}{\scriptsize{$\begin{array}{c} \text{iden } blong \\ gblong = G^{blong} \end{array}$}} 
      \end{tabular}
    }{} 
    \nextlevel[0.75]  
    \action*{\parbox{120pt}{\footnotesize{
          $\begin{array}[c]{l}
             \text{generates $bo$}\\
             \text{$gbo = G^{bo}$}\\
           \end{array}$
         }}}{B}
    \nextlevel[3.5]  
    \mess{\messLabel{$gblong$, $gbo$}}{B}{M}
    \nextlevel[0.4]  
    \action*{\parbox{120pt}{\footnotesize{
          $\begin{array}[c]{l}
             \text{replaces $gbo$ with $gbm$}\\
           \end{array}$
         }}}{M}
    \nextlevel[2.5]  
    \mess{\messLabel{$gblong$, $gbm$}}{M}{A}
    \nextlevel[0.3]  
    \condition{\messLabel{{\footnotesize off-band verification of $gblong$}}}{B,M,A}
    \nextlevel[1.2]  
    \action*{\parbox{200pt}{\footnotesize{
          $\begin{array}[c]{l}
             \text{generates $ae_1, ao$}\\
             \text{$gae_1 = G^{ae_1}$}\\
             \text{$gao = G^{ao}$}\\
             \text{$amaster = $HASH($gblong^{along}, gbm^{along}, gblong^{ae_1}$)}\\
             \text{$arkba_1$, $ackba_1$ = HKDF($amaster$)}\\
             \text{$acmba_1$ = HKDF(MAC($ackba_1$))}\\
           \end{array}$
         }}}{A}

    \nextlevel[6.5]  
    \action*{
      \parbox{170pt}{\footnotesize{
        \text{
          $\begin{array}[c]{l}
            \text{// Megolm session} \\
            \text{sender key $ska$} \\
            \text{$gska$ = $G^{ska}$} \\
            \text{generates $ra_1$, $ta_1$} \\
            \text{$gta_1$ = $G^{ta_1}$} \\
            \text{$xska$ = ENC($acmba_1, gska$)} \\
            \text{$xra_1$ = ENC($acmba_1, ra_1$)} \\
             \text{generates $ma_1$}\\
             \text{$aes\_a_1,hmac\_a_1,aesd\_a_1$ = HKDF($ra_1$)} \\
             \text{$xma_1 = $ AEAD\_ENC($aes\_a_1, ma_1, aesa\_d_1$)} \\
             \text{$xmasig_1 = $ SIGN($ska$, $xma_1$)}
          \end{array}$
        }}}}{A}
    \nextlevel[12]  
    \mess{\messLabel{$galong, gao$}}{A}{M}
    \mess{\messLabel{$galong, gao$}}{M}{B}
    \nextlevel[0.4]  
    \condition{\messLabel{{\footnotesize off-band verification of $galong$}}}{B,M,A}
    \nextlevel[2.25]  
    \mess{\messLabel{$xska, xra_1, xma_1, xmasig_1, gta_1$}}{A}{M}
    \mess{\messLabel{$xska, xra_1, xma_1, xmasig_1, gta_1$}}{M}{B}
    \nextlevel[0.4]  
    \action*{\parbox{135pt}{\footnotesize{
          $\begin{array}[c]{l}
             \text{Computes Olm, Megolm session info}\\
             \text{Decrypts $ma_1$}\\
             \text{\ldots Continues Megolm interactions}\\
           \end{array}$
         }}}{B}
    \nextlevel[4.5]  
    \mess{\messLabel{\textbf{Leaks} $blong$}}{B}{M}
    \nextlevel[0.4]  
    \action*{\parbox{200pt}{\footnotesize{
          $\begin{array}[c]{l}
             \text{$bmaster$ = HASH($blong^{galong}$, $gblong^{gbm}$, $gae_1^{blong}$)}\\
             \text{$brkba_1$, $bckba_1$ = HKDF($bmaster$)}\\
             \text{$bcmba_1$ = HKDF(MAC($bckba_1$))}\\
             \text{$ska$ = ENC($bcmba_1, xska$)} \\
             \text{$ra_1$ = ENC($bcmba_1, xra_1$)} \\
             \text{$aes\_b_1,hmac\_b_1,aesd\_b_1$ = HKDF($ra_1$)} \\
             \text{$ma_1 = $ AEAD\_DEC($ma_1$, $bes\_a_1, besa\_d_1$)} \\
             \text{\textbf{Obtains} $ra_1, ska, ma_1$}
           \end{array}$
         }}}{M}


   \end{msc}
   \vspace{-5pt}
   \caption{Post-compromise attacker sequence on unsigned Olm \& Megolm composition, as described in \ref{sub:OlmMegolm PCS attack}}
  \label{fig:megolm-pcs-vis}
\end{figure*}

\subsubsection{Signal identity key compromise does not compromise the reliant Sender Keys session}%
Because Signal maintains forward secrecy, the Sender Keys protocol (which employs Signal over Olm as the P2P channel) is \textit{not} vulnerable to the aforementioned attack. We prove this for the active and passive attacker case.

\subsection{Impact of Megolm post-compromise security limitations}%
\label{sub:Impact of Megolm post-compromise security limitations}
In brief, the attack described in \ref{sub:OlmMegolm PCS attack} implies compromising an Olm identity key of a peer that chooses to not sign its pre-keys allows the attacker to compromise all Megolm sessions of peers that initiate Olm sessions with said peer. To understand the impact of this limitation, we need to precisely understand how Olm sessions are handled in respect to Megolm sessions. We observe this detail is not properly specified in the Olm and Megolm specifications \cite{megolm-docs} \cite{olm-docs}. We confirmed two key details with the Matrix cryptography team regarding how Olm works in respect to Megolm. First, Olm identity keys are similar to Signal identity keys, i.e they last as long as the device the user is logged in on does. Second, a single Olm identity key is used for the connections with multiple different Olm peers. 

Thus, based on this detail, we can draw larger conclusions of how the compromise of an Olm identity key (in the case Olm pre-keys aren't signed) impacts Megolm over multiple epochs. \\

\noindent~$\bullet$~\textit{Only the first Megolm session sent to a peer with a compromised Olm identity key will be compromised by an active attacker}. Because Olm \textit{ratchets} forward (and thus is self-healing), we know only the \textit{first} Megolm session sent to a peer whose Olm identity key is compromised will be compromisable by an active attacker. \\

\noindent~$\bullet$~\textit{Any peer that initiates a Megolm connection with a compromised peer could have their first Megolm session compromised by an active attacker}. This is due to Olm identity keys being shared among multiple Olm sessions. \\

\noindent~$\bullet$~\textit{Newly joining Megolm peers could have their first Megolm session compromised by an active attacker; Megolm peers already in the group will not have their Megolm sessions compromised in future Epochs}. When a user joins a Megolm room, they must \textit{initiate} Olm sessions with all Megolm peers, potentially including a peer whose Olm identity key is compromised. However, because Olm is self-healing, we see new users will only have the messages of their first Megolm epoch compromised. Informally, we observe Olm's self-healing property \textit{propagates} to Megolm in a way.

Regarding the propagation of the aforementioned limitations to the Matrix protocol, we confirm the overarching Matrix protocol remains unaffected due to mandating the signing of Olm pre-keys \cite{matrix2024client}. 

\section{Related Work}

In this section we review broader related work other than the Matrix-related work which we reviewed in Section \ref{sec:bk_sec}.

\subsection{Computer-Aided Protocol Verification}%
\label{sub:Cryptographic Protocol Analysis Tools}


\textbf{Symbolic Cryptographic Analysis}. Most closely related to \vp, many other \textit{automated} symbolic protocol analysis tools exist. Most prevalent are \tamarin, another symbolic cryptanalysis tool \cite{Basin_Cremers_Dreier_Sasse_2022} and \pv \cite{blanchet2016modeling}.
\tamarin has been used to analyze Apple iMessage \cite{applePQ3}, TLS 1.3 \cite{Cremer_2017}, the EMV standard \cite{Basin_Sasse_Toro-Pozo_2021}, and The Noise protocol suite \cite{Girol_Hirschi_Sasse_Jackson}. Similarly \pv has been used to analyze Signal \cite{nadim-automated} and TLS 1.3 \cite{Bhargavan_Blanchet_Kobeissi_2017}. Other symbolic cryptanalysis tools such as \textsc{Scyther} \cite{Cremers_2008} and \textsc{AVISPA} \cite{Viganò_2006} have also been historically used for this purpose.

In comparison, \textit{manual} symbolic cryptanalysis has been theoretically feasible \cite{Bhargavan_Fournet_Gordon} but seldom employed until recent advances in proof-oriented programming languages. In particular, F*, a general-purpose proof-oriented language with support for dependent types proving, refinement times, executable semantics extractions, and verified implementation extractions to F\#, C, or Wasm \cite{Swamy_Martínez_Rastogi}, has seen usage for this purpose. Additionally, other dependent types-based theorem provers such as Rocq \cite{bertot_casteran_2004} and Agda \cite{norell_2009} are suited for this purpose, as they both offer executable semantics and implementation extractions
\footnote{Note, Rocq and Agda have different type theories: Rocq is based on the Calculus of Constructions (CoC) while Agda is based on Martin-L\"of Type Theory. This difference results in different approaches to constructing proofs. It is entirely unexplored how this difference affects the formulation of cryptographic protocols via the methodology described in \cite{Bhargavan_Fournet_Gordon}.}. The theoretical framework for verifying protocols symbolically using dependent-types introduced in \cite{Bhargavan_Fournet_Gordon} has been formalized in F* and was used to analyze Signal \cite{Bhargavan_dy} and 
later the Message Layer Security protocol \cite{Bhargavan_Beurdouche_Naldurg}.\footnote{Note that F* and dependent-types theorem provers are incredibly \textit{general} and are expressive enough to capture all of mathematics. Interestingly, F* has been additionally used to construct verified cryptographic libraries \cite{Beurdouche_2017} and to reason about group key agreement \cite{Beurdouche_thesis}.} 
However, constructing such formalizations in a dependent types prover such as F* is generally much more time consuming than constructing models in a symbolic cryptanalysis tool.



\textbf{Computational Cryptographic Analysis}. The most prevalent \textit{automated} computational cryptanalysis tool is \textsc{CryptoVerif}, which has been used to verify Signal \cite{nadim-automated}, SSH \cite{Cade_Blanchet_Paris-Rocquencourt}, TLS 1.3 \cite{Bhargavan_Blanchet_Kobeissi_2017}, and WireGuard \cite{Lipp_Blanchet_Bhargavan_2019}. \textsc{Easycrypt}, alternatively, features manual tactic-based reasoning with automated methods for simple lemmas \cite{Pereira}, and has been used to verify zero-knowledge protocols \cite{Firsov_Unruh_2023} and TLS \cite{Bhargavan_2014}. 
In general, computational cryptographic analysis tools make assumptions about the complexity of cryptographic primitives and reason about cryptosystems accordingly. We observe that there exists no constructive mechanization of the explicit computational complexity of cryptographic primitives. In fact, in stark contrast to other areas of math, complexity theory has seen virtually no mechanization in dependent types-based theorem provers \cite{Kunze_2021} \cite{leanprover2024}.

\subsection{Group Messaging Protocols With Key Agreement}%
\label{sub:Group Messaging Protocols With Key Agreement}

While the Megolm and Sender Keys protocols choose to approach group messaging without key agreement, there exists a large body of work analyzing protocols \textit{with} group key agreement. Group key agreement protocols have a long history of formal \cite{Poettering_Rösler_Schwenk_Stebila_2021,gdh_ccs96,tgdh_tissec2004} and empirical \cite{Amir_Kim_Nita-Rotaru} analysis. Recently, asynchronous ratcheting trees have been introduced as an effective method to achieve post-compromise security \cite{Cohn-Gordon_Cremers_Garratt_Millican_Milner_2018}, forming the basis for the development of Message Layer Security \cite{rfc9420}. Group messaging protocols based on asynchronous ratchet trees have been analyzed and verified with \textsc{Tamarin} \cite{Cremers_Jacomme_Lukert_2023} and F* \cite{Bhargavan_Beurdouche_Naldurg}.

At the time of writing, the Matrix foundation intends to phase out Olm and Megolm in favor of Message Layer Security \cite{AreWeMLSYet} \cite{MLSblog} but has yet to do so. In addition to better post-compromise security guarantees \cite{Cohn-Gordon_Cremers_Garratt_Millican_Milner_2018}, Message Layer Security has been shown to be more performant than Olm and Megolm in practice \cite{Chathi2022MLSComparison}.

\section{Conclusion}%
\label{sec:Conclusion}

This paper has presented a comprehensive formal, symbolic analysis of the Olm and Megolm protocols within the Matrix cryptographic suite. Utilizing \pv, we constructed precise models and conducted automated symbolic analysis to verify key security properties, including confidentiality, authentication, forward secrecy, and post-compromise security for Olm and Megolm. Our approach enabled us to precisely and automatically reproduce known vulnerabilities and attacks. 

In addition, through our comparative analysis with the Signal and Sender Keys protocols, we highlighted critical differences in post-compromise security, emphasizing the need for ongoing scrutiny and improvement of secure messaging protocols. We endorse the Matrix protocol specification's current decision for Olm pre-keys to be signed, as our models confirm that when Olm pre-keys are signed the security guarantees of Olm and Megolm are comparable to Signal and Sender Keys. 

As for future work, we plan to round out our symbolic analysis of the Matrix cryptographic suite with a formalization in the dependent types-based symbolic analysis framework DY*, which is implemented in F* \cite{Bhargavan_dy}. This would allow us to construct secrecy and authentication proofs that are parameterized over the number of Olm ratchet iterations and Megolm epoch iterations, as well as extract executable semantics to more effectively reason about the implementations. 

Overall, our findings underscore the importance of formal, mechanized proofs in ensuring the robustness of cryptographic systems, particularly in the context of federated and decentralized communication platforms like Matrix. We hope our work motivates further advancement of cryptographic verification tooling and the continued analysis and scrutiny of the protocols of which we all rely upon.



\bibliographystyle{IEEEtranS}
\bibliography{refs}

\section{Appendix}%
\label{sec:Appendix}

\subsection{Megolm Ratchet Definition}%
\label{sub:Megolm Ratchet Definition}

We briefly describe the Megolm ratchet scheme as described in the Megolm specification \cite{megolm-docs}. The Megolm ratchet $R_i$ consists of four components, $R_{i,j}$ for $j \in \{0,1,2,3\}$, each 256 bits in this implementation. Initialization is done with cryptographically-secure random data and advances as follows:
\[
\begin{aligned}
R_{i,0} &= \begin{cases}
  H_0\left(R_{2^{24}(n-1),0}\right) & \text{if } i = 2^{24}n \\
  R_{i-1,0} & \text{otherwise}
\end{cases} \\
R_{i,1} &= \begin{cases}
  H_1\left(R_{2^{24}(n-1),0}\right) & \text{if } i = 2^{24}n \\
  H_1\left(R_{2^{16}(m-1),1}\right) & \text{if } i = 2^{16}m \\
  R_{i-1,1} & \text{otherwise}
\end{cases} \\
R_{i,2} &= \begin{cases}
  H_2\left(R_{2^{24}(n-1),0}\right) & \text{if } i = 2^{24}n \\
  H_2\left(R_{2^{16}(m-1),1}\right) & \text{if } i = 2^{16}m \\
  H_2\left(R_{2^8(p-1),2}\right) & \text{if } i = 2^8p \\
  R_{i-1,2} & \text{otherwise}
\end{cases} \\
R_{i,3} &= \begin{cases}
  H_3\left(R_{2^{24}(n-1),0}\right) & \text{if } i = 2^{24}n \\
  H_3\left(R_{2^{16}(m-1),1}\right) & \text{if } i = 2^{16}m \\
  H_3\left(R_{2^8(p-1),2}\right) & \text{if } i = 2^8p \\
  H_3\left(R_{i-1,3}\right) & \text{otherwise}
\end{cases}
\end{aligned}
\]

Here, $H_0$, $H_1$, $H_2$, and $H_3$ are distinct hash functions. In summary:

\begin{itemize}
  \item Every $2^8$ iterations, $R_{i,3}$ is reseeded from $R_{i,2}$.
  \item Every $2^{16}$ iterations, $R_{i,2}$ and $R_{i,3}$ are reseeded from $R_{i,1}$.
  \item Every $2^{24}$ iterations, $R_{i,1}$, $R_{i,2}$, and $R_{i,3}$ are reseeded from $R_{i,0}$.
\end{itemize}

The complete ratchet value $R_i$ is hashed to generate encryption keys. This design allows arbitrary forward ratchet advancement, requiring at most 1020 hash computations, and therefore minimizing cost. A client can decrypt chat history from the earliest known ratchet value but cannot decrypt prior history without reversing the hash.


\section{Protocol Handshakes}%
\label{sec:Protocol Handshakes}

\definecolor{gris}{gray}{0.85}
\begin{figure*}[ht!]
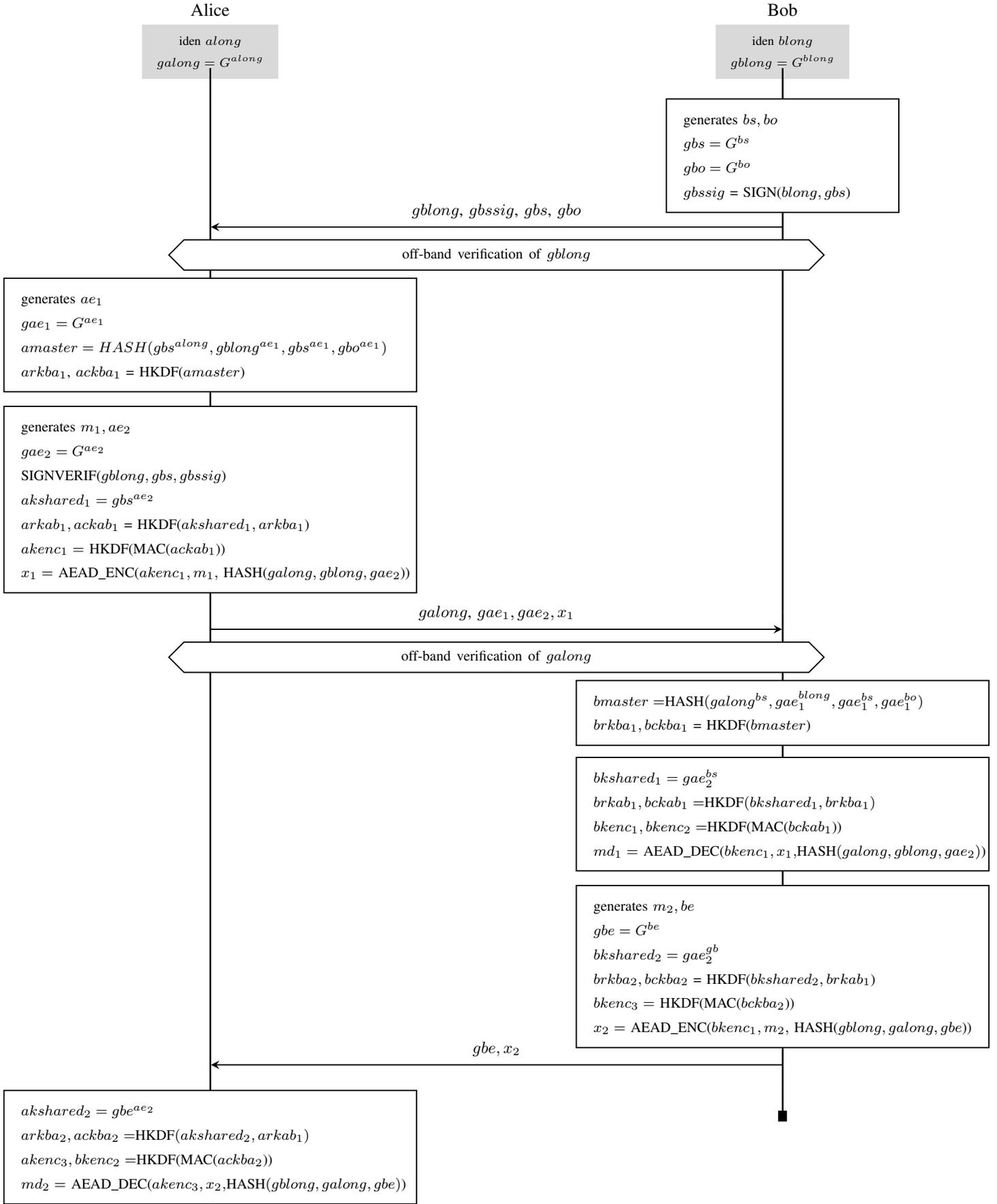

  \centering
  \setmsckeyword{}
  \drawframe{no}    
  \begin{msc}[
    /msc/title top distance=0cm,
    /msc/first level height=.1cm,
    /msc/last level height=0.4cm,
    /msc/head height=0cm,
    /msc/instance width=0cm,
    /msc/head top distance=0.5cm,
    /msc/foot distance=-0.0cm,
    /msc/instance width=0cm,
    /msc/condition height=0.2cm
    ]{}
    \setlength{\instwidth}{0\mscunit} 
    \setlength{\instdist}{11cm}  

    \declinst{A}{              
      \begin{tabular}[c]{c}
        Alice \\ 
        \colorbox{gris}{\scriptsize{$\begin{array}{c} \text{iden }along \\ galong = G^{along}\end{array}$}} 
      \end{tabular}
    }{}
    \declinst{B}{              
      \begin{tabular}[c]{c}
        Bob \\ 
        \colorbox{gris}{\scriptsize{$\begin{array}{c} \text{iden } blong \\ gblong = G^{blong} \end{array}$}} 
      \end{tabular}
    }{} 

    \nextlevel[1]
    \action*{\parbox{120pt}{\footnotesize{
          $\begin{array}[c]{l}
             \text{generates $bs, bo$}\\
             \text{$gbs = G^{bs}$}\\
             \text{$gbo = G^{bo}$}\\
             \text{$gbssig$ = SIGN($blong, gbs$)}\\
           \end{array}$
         }}}{B}
    \nextlevel[5]
    \mess{\messLabel{$gblong$, $gbssig$, $gbs$, $gbo$}}{B}{A}
    \nextlevel[0.5]
    \condition{\messLabel{{\footnotesize off-band verification of $gblong$}}}{B,A}
    \nextlevel[1.5]

    \action*{\parbox{220pt}{\footnotesize{
          $\begin{array}[c]{l}
             \text{generates $ae_1$}\\
             \text{$gae_1 = G^{ae_1}$}\\
             \text{$amaster = HASH(gbs^{along}, gblong^{ae_1}, gbs^{ae_1}, gbo^{ae_1})$}\\
             \text{$arkba_1$, $ackba_1$ = HKDF($amaster$)}\\
           \end{array}$
         }}}{A}
    \nextlevel[5]
    \action*{\parbox{220pt}{\footnotesize{
          $\begin{array}[c]{l}
             \text{generates $m_1, ae_2$}\\
             \text{$gae_2 = G^{ae_2}$}\\
             \text{SIGNVERIF($gblong, gbs, gbssig$)}\\
             \text{$akshared_1 = gbs^{ae_2}$}\\
             \text{$arkab_1, ackab_1$ = HKDF($akshared_1, arkba_1$)}\\
             \text{$akenc_1 = $ HKDF(MAC($ackab_1$))}\\
             \text{$x_1 = $ AEAD\_ENC($akenc_1,m_1,$ HASH($galong, gblong, gae_2$))}\\
           \end{array}$
         }}}{A}
    \nextlevel[8.75]
    \mess{\messLabel{$galong$, $gae_1,gae_2,x_1$}}{A}{B}
    \nextlevel[0.5]
    \condition{\messLabel{{\footnotesize off-band verification of $galong$}}}{B,A}
    \nextlevel[1.5]
    \action*{\parbox{220pt}{\footnotesize{
       $\begin{array}[c]{l}
         \text{$bmaster = $HASH$(galong^{bs}, gae_1^{blong}, gae_1^{bs}, gae_1^{bo})$} \\
         \text{$brkba_1, bckba_1$ = HKDF($bmaster$)}
        \end{array}$
       }}}{B}
    \nextlevel[3]
    \action*{\parbox{220pt}{\footnotesize{
       $\begin{array}[c]{l}
         \text{$bkshared_1 = gae_2^{bs}$} \\
         \text{$brkab_1, bckab_1 = $HKDF$(bkshared_1,brkba_1)$} \\
         \text{$bkenc_1, bkenc_2 = $HKDF(MAC($bckab_1$))} \\
         \text{$md_1 = $ AEAD\_DEC($bkenc_1, x_1, $HASH$(galong, gblong, gae_2)$)}
        \end{array}$
       }}}{B}
    \nextlevel[5]
    \action*{\parbox{220pt}{\footnotesize{
          $\begin{array}[c]{l}
             \text{generates $m_2, be$}\\
             \text{$gbe = G^{be}$}\\
             \text{$bkshared_2 = gae_2^{gb}$}\\
             \text{$brkba_2, bckba_2$ = HKDF($bkshared_2, brkab_1$)}\\
             \text{$bkenc_3 = $ HKDF(MAC($bckba_2$))}\\
             \text{$x_2 = $ AEAD\_ENC($bkenc_1,m_2,$ HASH($gblong, galong, gbe$))}\\
           \end{array}$
         }}}{B}
  \nextlevel[7]
  \mess{\messLabel{$gbe, x_2$}}{B}{A}
  \nextlevel]
    \action*{\parbox{220pt}{\footnotesize{
       $\begin{array}[c]{l}
         \text{$akshared_2 = gbe^{ae_2}$} \\
         \text{$arkba_2, ackba_2 = $HKDF$(akshared_2,arkab_1)$} \\
         \text{$akenc_3, bkenc_2 = $HKDF(MAC($ackba_2$))} \\
         \text{$md_2 = $ AEAD\_DEC($akenc_3, x_2, $HASH$(gblong, galong, gbe)$)}
        \end{array}$
       }}}{A}

   \end{msc}
   \vspace{-15pt}
  \caption{Complete Signal Protocol}
  \label{fig:signal}
\end{figure*}

\end{document}